\documentclass[conference]{IEEEtran}
\ifCLASSINFOpdf
\else
\fi
\usepackage[cmex10]{amsmath}
\ifCLASSOPTIONcompsoc
  \usepackage[caption=false,font=normalsize,labelfont=sf,textfont=sf]{subfig}
\else
  \usepackage[caption=false,font=footnotesize]{subfig}
\fi
\newcommand{\matr}[1]{\mathbf{#1}}
\DeclareMathOperator{\diag}{diag}

\DeclareMathOperator{\expect}{\mathrm{E}}

\usepackage{pgfplots}
\pgfplotsset{compat=newest}
\newlength\figureheight	
\newlength\figurewidth

\newcommand{\columnplot}{
\setlength\figureheight{0.25\textwidth}
\setlength\figurewidth{0.38\textwidth}
}	

\usetikzlibrary{shapes,arrows}

\usetikzlibrary{external} 
\tikzexternaldisable

\usepackage[exponent-product = \cdot]{siunitx}

\begin{document}
%
\title{Performance Evaluation of LOS MIMO Systems under the Influence of Phase Noise}

\author{\IEEEauthorblockN{Tim H{\"a}lsig and Berthold Lankl}
\IEEEauthorblockA{Institute for Communications Engineering\\
Universit{\"a}t der Bundeswehr M{\"u}nchen, Germany\\
Email: tim.haelsig@unibw.de}
}


%


\maketitle

\begin{abstract}
To deal with the ever increasing data rate demand in wireless systems, suitable backhaul solutions have to be found. One option is to use LOS MIMO systems at mmWave frequencies, as high bandwidths and spectral efficiencies will be able to fulfill this requirement. Several investigations have shown, that phase noise can decimate the efficiency of such systems immensely  and needs to be adequately dealt with. In this work we investigate the system level performance of LOS MIMO systems subject to this impairment and examine the influence for various practical phase noise values. Furthermore, it is shown how a simple phase noise compensation algorithm can be used to enhance the performance of these systems in different scenarios.
\end{abstract}


\let\thefootnote\relax\footnotetext{This work has been supported by the German Research Foundation~(DFG) in the framework of priority program SPP~1655 "Wireless Ultra High Data Rate Communication for Mobile Internet Access".}

%
\IEEEpeerreviewmaketitle

\section{Introduction}

The high level of densification in wireless networks, which will increase even further if millimeter wave systems are used for future cellular communication, demands highly efficient and flexible backhaul solutions. One of the best solutions for such networks is wireless backhauling as it is very flexible and can be very cost efficiently deployed.
To cope with the high data rates that are required for backhauling, it will be necessary to design systems with a very high spectral efficiency. Multiple-input multiple-output~(MIMO) has become the enabling technique to build such efficient systems. The combination of MIMO techniques and the huge amount of available bandwidth at millimeter wave frequencies, e.g., up to four $\SI{2.16}{\giga\hertz}$ channels at \SI{60}{\giga\hertz}, thus seems as an appealing option to achieve the required data rates for future backhauling.

As millimeter waves experience a high path loss it will be almost mandatory, especially for backhaul systems, to operate in a line-of-sight~(LOS) scenario. For such a setup, the often well justified Rayleigh channel assumption, which results in high spectral efficiencies for MIMO systems, does not apply anymore. Nevertheless, it was shown in \cite{Bohagen2007} that high spectral efficiency and spatial multiplexing is possible for LOS MIMO systems, by carefully adjusting the antenna arrangement. Experimental validation of this concept was for example done in \cite{Sheldon2009}, where the authors spatially multiplexed four streams over a distance of \SI{5}{\metre} at \SI{60}{\giga\hertz}.

In practice the transmitted and received baseband signal will not only be influenced by the wireless channel, but also by the characteristic of the transmit and receive radio~frequency~(RF) components. It is thus of great interest to investigate how much deterioration due to RF impairments is tolerable, and if it can be efficiently compensated. There exists a tremendous amount of work on this topic, an overview of several impairments can, e.g., be found in \cite{Schenk2008}. Recently, there has also been a growing interest in the influence of RF impairments on MIMO systems. Information theoretic analyses of MIMO systems with hardware impairments in \cite{Bjornson2013,Zhang2014} revealed that although there is a finite capacity ceiling at high SNR, the capacity still scales if the number of antennas is increased both on the transmitter and receiver side, and thus spatial multiplexing is possible. Phase noise as one specific deterioration has been investigated in \cite{Durisi2013,Durisi2014}. In the first paper the authors concluded that when the phase noise processes at the antennas are independent, memoryless in time and uniformly distributed, no multiplexing gain is to be expected. The second paper introduces capacity bounds for the case of a common oscillator setup.
Finally, in \cite{Mehrpouyan2012} the authors developed a scheme to estimate the MIMO channel and to track phase noise, which is based on a decision-directed extended Kalman filter. System level simulations show that the proposed method can significantly improve the BER of MIMO systems, even for high phase noise values.

In this work we will perform a similar system level study on LOS MIMO systems subject to phase noise, where we will investigate the difference between the common and individual oscillator setup, examine the impact of different practical oscillator models and show a simple phase noise compensation scheme, which works well for low levels of phase noise.

Let $(\cdot)^*$, $(\cdot)^T$, $(\cdot)^H$ denote conjugate, transpose and conjugate transpose, respectively, while $\expect\left[\cdot\right]$ denotes the expectation operator. Furthermore, boldface small letters, e.g., $\matr{x}$, are used for vectors while boldface capital letters, e.g., $\matr{X}$, are used for matrices. The $m\times m$ identity matrix is given as $\matr{I}_m$ and $\diag\left(\matr{x}\right)$ denotes the diagonal matrix where the diagonal elements are given in vector $\matr{x}$.

\section{System Model}

Assume that the received signal of an equivalent discrete baseband MIMO system is given by
\begin{equation}
\matr{y}(n) = \matr{\Theta}_\text{Rx}(n)\matr{H}\matr{\Theta}_\text{Tx}(n)\matr{x}(n) + \matr{w}(n) \label{eq:sysmodel}
\end{equation}
where $\matr{x}(n)=\left[x_1(n),\ldots,x_{N_\text{Tx}}(n)\right]^T$ is the $n$th transmitted symbol vector with sampling rate of $1/T_s$, $\matr{y}(n)=\left[y_1(n),\ldots,y_{N_\text{Rx}}(n)\right]^T$ is the $n$th received symbol vector, and $N_\text{Tx}$ and $N_\text{Rx}$ being the number of transmit and receive antennas, respectively. We will in the following only consider systems where the number of antennas is the same on both sides of the link, i.e., $N_\text{Tx}=N_\text{Rx}=N$. Furthermore, let $\matr{w}(n)$ be a complex additive noise vector with complex Gaussian distribution, i.e., $\matr{w}(n)\sim\mathcal{C}\mathcal{N}(0,\sigma^2_w\matr{I}_{N_\text{Rx}})$, and let $\matr{H}$ be the channel gain matrix of dimensions $N_\text{Rx}\times N_\text{Tx}$. 

The channel matrix $\matr{H}$ will in the following consist of two parts given by
\begin{equation}
\matr{H} = \sqrt{\frac{K}{1+K}}\matr{H}_\text{LOS} + \sqrt{\frac{1}{1+K}}\matr{H}_\text{NLOS} 
\end{equation}
where $K$ is the Rician factor used to weigh the impact of each component. We will assume $\matr{H}_\text{LOS}$ to be unitary, i.e., antenna arrangement designed according to \cite{Bohagen2007}, while $\matr{H}_\text{NLOS}$ is assumed to have i.i.d. complex Gaussian entries with zero mean and unit variance. Note that for millimeter wave communication in general and in a backhaul scenario specifically, highly directional antennas will have to be used in order to combat high path losses. For a LOS setup it seems thus reasonable to assume fairly high $K$~factors, i.e., above \SI{10}{\decibel} see \cite{Muhi-Eldeen2010}.

The two matrices $\matr{\Theta}_\text{Tx}(n)$ and $\matr{\Theta}_\text{Rx}(n)$ in Equation~\eqref{eq:sysmodel} represent the phase noise processes at transmitter and receiver, respectively. In the most general setup they are diagonal matrices, i.e.,
\begin{equation}
\matr{\Theta}_\text{Tx}(n) = \diag\left(\left[e^{j\theta_{1}^{\text{Tx}}(n)},\ldots,e^{j\theta_{N_\text{Tx}}^{\text{Tx}}(n)}\right]^T\right) \notag
\end{equation}
\begin{equation}
\matr{\Theta}_\text{Rx}(n) = \diag\left(\left[e^{j\theta_{1}^{\text{Rx}}(n)},\ldots,e^{j\theta_{N_\text{Rx}}^{\text{Rx}}(n)}\right]^T\right) \text{,}\notag
\end{equation}
where for example $\theta_{i}^\text{Tx}(n)$ denotes the phase noise sample at time $n$ for the $i$th transmit antenna, whose generation will be explained in the next section.

When modeling each phase noise process for each transmit and each receive antenna individually, this model corresponds to the case where each separate antenna has its individual oscillator. On the other hand, when only one phase noise process is modeled at Tx and Rx, e.g., $\theta_{1}^\text{Tx}(n)=...=\theta_{N_\text{Tx}}^\text{Tx}(n)$, this model corresponds to the common oscillator setup and the system model can be simplified to 
\begin{equation}
\matr{y}(n) = e^{j\theta_{i}^{\text{Rx}}(n)}e^{j\theta_{i}^{\text{Tx}}(n)}\cdot\matr{H}\matr{x}(n) + \matr{w}(n)
\end{equation}
which is generally assumed to be favorable compared to the individual oscillator setup. Nevertheless, it is of interest to investigate both cases as for optimal LOS MIMO system design, the antennas can be very widely spaced, e.g., \SI{60}{\giga\hertz}, link distance of \SI{100}{\metre} and $N_\text{Tx}=2$ leads to a spacing of \SI{0.5}{\metre}, and carrier distribution over such long lines will be a very challenging circuit design task.

\section{Phase Noise Models}
Phase noise~(PN) generally arises due to the fact that a practical oscillator can never generate a perfect sinusoid. In the equivalent baseband model this translates to a slowly varying phase shift from symbol to symbol. The impact has been extensively studied and various models exist for the characterization of the process. We will, in this work, focus on the two most common ones: Wiener phase noise and stationary phase noise, which will be explained in the next two sections. It has been observed in \cite{Khanzadi2014} that the white noise regions far from the carrier have a significant impact for wideband systems, which will be taken into account when generating the phase noise processes.

\subsection{Wiener Phase Noise}
The Wiener process is a widely employed modeling approach for the phase noise of a free running oscillator, c.f. \cite{Demir2000}. In discrete time it can be defined as
\begin{equation}
\theta_i(n) = \theta_i(n-1) + \Delta_i(n)
\end{equation}
where the phase change at time $n$ for Tx/Rx antenna $i$ is modeled as a Gaussian process following $\Delta_i(n)\sim\mathcal{N}(0,\sigma^2_{\Delta_i})$. The variance of the phase change $\sigma^2_{\Delta_i}$ will in the following be the same across all antennas at transmitter and receiver. It can be directly related to the sampling rate $1/T_s$ and the \SI{3}{\decibel}~bandwidth $\beta_i$ of an oscillator's phase noise spectrum with $\sigma^2_{\Delta_i}=4\pi \beta_i T_s$. 

For example, we will evaluate $\sigma^2_{\Delta_i}= \SI{e-4}{\radian^2}$ and $\sigma^2_{\Delta_i}=\SI{e-5}{\radian^2}$ with a sampling time of $T_s = \SI{e-9}{\second}$, equivalent to $\beta_i \approx \SI{8}{\kilo\hertz}$ and $\beta_i \approx \SI{0.8}{\kilo\hertz}$, respectively. These can further be translated to phase noise levels of approximately \SI{-86}{\decibel c\per\hertz} and \SI{-96}{\decibel c\per\hertz} at \SI{1}{\mega\hertz} offset, which are reasonable values, c.f. \cite{Rappaport2011}. Generally, the phase change from symbol to symbol will indeed be very small, e.g., below \SI{1.5}{\degree} for $\beta_i \approx \SI{8}{\kilo\hertz}$.

\subsection{Stationary Phase Noise}
For a phase locked oscillator the phase noise is often modeled as a stationary process given by
\begin{equation}
\theta_i(n) = \theta_{i,0} + \phi_i(n)
\end{equation}
where $\theta_{i,0}$ is a constant phase offset and $\phi_i(n)$ is a zero mean, WSS, colored Gaussian process, which is generated by applying a digital filter, with a desired power spectral density, to a zero mean and unit variance Gaussian random variable \cite{Kasdin1995}.

As reference, we will use filters with the power spectral densities measured in \cite{Reynolds2006} and \cite{Dancila2014}, which have phase noise levels at \SI{1}{\mega\hertz} offset of \SI{-85}{\decibel c\per\hertz} and \SI{-115}{\decibel c\per\hertz}, respectively.

\section{Transmission Scheme}

We will use a frame based transmission scheme, similar to the one used in \cite{Mehrpouyan2012}, where the data frame of each antenna is preceded by a training sequence of $L_t$ symbols, which can be used for channel estimation. The overall length of the frame will be $L_f = L_t+L_d$ symbols where $L_d$ is the number of data symbols in the frame. It is further assumed that the NLOS part of the channel $\matr{H}_\text{NLOS}$ does not change during the frame.

In the receiver a zero~forcing~(ZF) algorithm will be employed, given by
\begin{equation}
\hat{\matr{x}}(n) = \hat{\matr{H}}^{\dagger}\matr{y}(n)
\end{equation}
with $\hat{\matr{H}}^{\dagger}$ being the pseudo-inverse of the channel matrix estimate gained from the training sequence. Note that for $K=\infty$ and perfect LOS MIMO antenna arrangement, this algorithm corresponds to the maximum~likelihood (ML) receiver. 

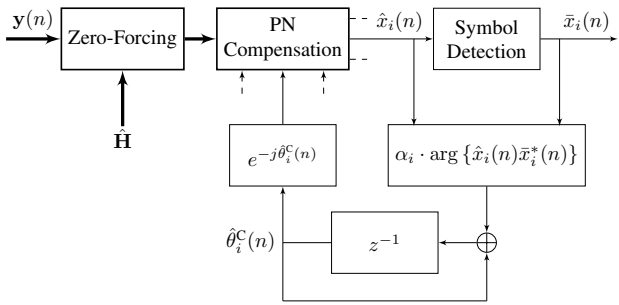
\begin{figure}[!t]
\centering
\resizebox{0.48\textwidth}{!}{\tikzstyle{block} = [draw, fill=white, rectangle,minimum height=3em, minimum width=5.2em]	
\tikzstyle{block_rot} = [draw, fill=white, rectangle,minimum height=5em, minimum width=3em]
\tikzstyle{sum} = [draw, fill=white, circle, node distance=1em,path picture={\draw[black](path picture bounding box.south) -- (path picture bounding box.north) (path picture bounding box.west) -- (path picture bounding box.east);}]

\tikzstyle{coord} = [coordinate]

\begin{tikzpicture}[auto, node distance=4cm,>=latex']
	\node [block,thick] (zf) {Zero-Forcing};
	\node [coord,left of=zf,node distance=2cm,label={}] (input) {};
	\draw [->,ultra thick] (input) -- node[pos=0.5] {$\matr{y}(n)$}(zf);
	
	\node [block,right of =zf,node distance=2.75cm,align=center,thick] (pncomp) {PN\\ Compensation};
	\draw [->,ultra thick] (zf) -- node[pos=0.5] {}(pncomp);
	
	\node [coord,below of=zf,node distance=1.5cm,label={},anchor=east] (Hest) {};
	\node [coord,below of=Hest,node distance=0.5cm,label={$\hat{\matr{H}}$},anchor=east] (Hname) {};
	\draw [->,ultra thick] (Hest) -- node[pos=0.5] {}(zf);
	
	\node [block,right of =pncomp,node distance=3.5cm,align=center] (det) { Symbol\\  Detection};
	\draw [->] (pncomp) -- node[pos=0.6] {$\hat{x}_i(n)$}(det);
	
	\node [coord,right of=det,node distance=2.25cm] (output) {};
	\draw [->] (det) -- node[pos=0.6] {$\bar{x}_i(n)$}(output);
	
	\node [block,below of =det,node distance=2cm,align=center] (arg) {$\alpha_i \cdot \arg\left\{\hat{x}_i(n)\bar{x}_i^*(n)\right\}$};
	
	\node [sum,below of =arg,node distance=1.5cm,align=center] (add) {};
	\node [block,left of =add,node distance=1.75cm,align=center] (z1) {$z^{-1}$};
	\draw [->] (arg) -- node[pos=0.6] {}(add);
	\draw [->] (add) -- node[pos=0.6] {}(z1);
	
	\node [block,left of =arg,node distance=3.5cm,align=center] (exp) {$e^{-j\hat{\theta}^\text{C}_i(n)}$};
	\node [coord,below of=exp,node distance=1.5cm] (expcon) {};
	\draw [-] (z1) -- node[pos=0.6] {}(expcon);
	\draw [->] (expcon) -- node[pos=0.6] {}(exp);
	
	\node [coord,below of=expcon,node distance=1cm] (expcon2) {};
	\node [coord,below of=add,node distance=1cm] (sumcon) {};
	\draw [-] (expcon) -- node[pos=0.6] {}(expcon2);
	\draw [-] (expcon2) -- node[pos=0.6] {}(sumcon);
	\draw [->] (sumcon) -- node[pos=0.6] {}(add);
	
	\draw [->] (exp) -- node[pos=0.6] {}(pncomp);
	
	\node [coord,left of=det,node distance=1.25cm] (argc1) {};
	\draw [->] (argc1) -- ++(0,-1.5);
	\node [coord,right of=det,node distance=1.25cm] (argc2) {};
	\draw [->] (argc2) -- ++(0,-1.5);
	
	\node [coord,left of=expcon,node distance=0.55cm,yshift=-3mm,label={$\hat{\theta}_i^\text{C}(n)$},anchor=north] (thname) {};
	
	\draw [-,dashed] (pncomp.east)++(0,0.35) -- ++(0.3,0);
	\draw [-,dashed] (pncomp.east)++(0,-0.35) -- ++(0.3,0);
	
	\draw [<-,dashed] (pncomp.south)++(0.7,0) -- ++(0,-0.5);
	\draw [<-,dashed] (pncomp.south)++(-0.7,0) -- ++(0,-0.5);
	
	
\end{tikzpicture}}
\caption{Zero-forcing receiver structure with phase noise compensation scheme based on a feedback loop.}
\label{fig:ZFreceiver}
\end{figure}

\subsection{Phase Noise Compensation}

It has been shown previously that compensating the impact of phase noise in a system can greatly improve its performance. It is easy to see that if perfect knowledge about the phase noise processes exists at the receiver, i.e., $\matr{\Theta}_\text{Tx}(n)$ and $\matr{\Theta}_\text{Rx}(n)$ are known, the influence could be totally compensated by premultiplying $\matr{y}(n)$ with the pseudo-inverse of the term $\matr{\Theta}_\text{Rx}(n)\matr{H}\matr{\Theta}_\text{Tx}(n)$.

In practice this will never be possible, also because the estimate $\hat{\matr{H}}$ will always be intrinsically influenced by the phase noise. We will thus use a combined phase noise matrix estimate $\hat{\matr{\Theta}}_\text{C}(n)$ which leads to the updated receiver algorithm given as
\begin{equation}
\hat{\matr{x}}(n) = \hat{\matr{\Theta}}_\text{C}(n)\hat{\matr{H}}^{\dagger}\matr{y}(n)
\end{equation}
where 
\begin{equation}
\hat{\matr{\Theta}}_\text{C}(n)=\diag\left(\left[e^{-j\hat{\theta}_{1}^\text{C}(n)},\ldots,e^{-j\hat{\theta}_{N}^\text{C}(n)}\right]^T\right) \text{.} \notag
\end{equation}
It is easily verified that for the common oscillator setup the matrix reduces to the factor $e^{-j\hat{\theta}_{i}^\text{C}(n)}$, where 
\begin{equation}
\theta_{i}^\text{C}(n)=\theta_{i}^\text{Tx}(n)+\theta_{i}^\text{Rx}(n) \label{eq:pncomp}
\end{equation}
holds and is the same for all parallel streams $i$. For the individual oscillator setup the impact is more complicated, it will usually depend on all phase noise processes in the transmitter and receiver. For compensation we will in this case nevertheless assume that the process $\theta_{i}^\text{C}(n)$ follows a slowly varying characteristic, but is independent for each stream $i$.

Given the above assumptions, easy phase noise compensation could be achieved by feeding back the phase difference between the ZF estimate $\hat{\matr{x}}(n)$ and the detected symbol denoted by $\bar{\matr{x}}(n)$. In particular for each stream $i$
\begin{equation}
\hat{\theta}_i^\text{C}(n) = \hat{\theta}_i^\text{C}(n-1) + \alpha_i\cdot\arg\left\{\hat{x}_i(n-1)\bar{x}_i^*(n-1)\right\} \label{eq:tracking}
\end{equation}
where $\alpha_i$ is a constant weighing the impact of each new phase estimate. For the common oscillator setup the compensation performance can further be improved by using the mean over the estimated phase noise processes $\hat{\theta}_i^\text{C}(n)$ across all streams $i$ for each time instant $n$. The full receiver structure can be seen in Fig.~\ref{fig:ZFreceiver}, where the dashed lines symbolize that the following parts are carried out in parallel for each stream $i$.

\begin{figure}[!t]
\centering
\tikzexternalenable
\tikzsetnextfilename{EVM_WienerStatComp}
\subfloat[]{\columnplot
%
%

\definecolor{mycolor1}{rgb}{1,0.47058823529,0.47058823529}

\begin{tikzpicture}

\begin{axis}[%
width=\figurewidth,
height=\figureheight,
scale only axis,
xmin=10,
xmax=35,
xlabel={SNR in dB},
xmajorgrids,
ymin=0,
ymax=0.5,
ylabel={EVM},
ymajorgrids,
legend style={at={(0.98,0.98)},anchor=north east,draw=black,fill=white,legend cell align=left,font=\footnotesize}
]
\addplot [color=black,dashed,thick]
  table[row sep=crcr]{10	0.316275362\\
12	0.251038922\\
14	0.199685607\\
16	0.158359427\\
18	0.125843099\\
20	0.100054347\\
22	0.079383316\\
24	0.063128098\\
26	0.050122222\\
28	0.039822088\\
30	0.031621005\\
32	0.02514122\\
34	0.019965213\\
36	0.015841431\\
38	0.012586521\\
40	0.009998994\\
};
\addlegendentry{No PN};

\addplot [color=blue,solid,dotted,thick,forget plot]
  table[row sep=crcr]{10	0.426535891\\
12	0.378118941\\
14	0.34360959\\
16	0.318153258\\
18	0.301102727\\
20	0.289291704\\
22	0.281731653\\
24	0.276537782\\
26	0.273178125\\
28	0.270904511\\
30	0.269535709\\
32	0.268651521\\
34	0.268177842\\
36	0.267766083\\
38	0.267542654\\
40	0.267401775\\
};

\addplot [color=blue,solid]
  table[row sep=crcr]{10	0.330113483\\
12	0.268411609\\
14	0.220943233\\
16	0.185355611\\
18	0.158049256\\
20	0.138223336\\
22	0.123653187\\
24	0.113716832\\
26	0.106934302\\
28	0.102284861\\
30	0.09931378\\
32	0.097379208\\
34	0.096113154\\
36	0.095315213\\
38	0.094831252\\
40	0.09445841\\
};
\addlegendentry{Wiener PN};

\addplot [color=blue,solid,forget plot]
  table[row sep=crcr]{10	0.452648061\\
12	0.408997979\\
14	0.379366757\\
16	0.358371625\\
18	0.344979704\\
20	0.335654209\\
22	0.330095184\\
24	0.326330842\\
26	0.324022359\\
28	0.322696243\\
30	0.321583576\\
32	0.321024601\\
34	0.320666371\\
36	0.320422206\\
38	0.320237723\\
40	0.320179686\\
};

\addplot [color=mycolor1,solid]
  table[row sep=crcr]{10	0.31684821\\
12	0.251729741\\
14	0.199646229\\
16	0.158804476\\
18	0.126477869\\
20	0.100597661\\
22	0.080148581\\
24	0.063856679\\
26	0.051197046\\
28	0.041257834\\
30	0.033289244\\
32	0.027186431\\
34	0.022539114\\
36	0.019017762\\
38	0.01640295\\
40	0.014547051\\
};
\addlegendentry{Stationary PN};

\addplot [color=mycolor1,solid,forget plot]
  table[row sep=crcr]{10	0.339336267\\
12	0.279602487\\
14	0.234559804\\
16	0.200657854\\
18	0.17625372\\
20	0.158579475\\
22	0.146408076\\
24	0.138272535\\
26	0.132834698\\
28	0.129131313\\
30	0.126871656\\
32	0.125434089\\
34	0.124462575\\
36	0.123888728\\
38	0.123489184\\
40	0.123228887\\
};

\node[label={\footnotesize Reynolds}] at (30,0.11) {};
\node[label={\footnotesize Dancila}] at (30,0.02) {};

\node[pin=90:{\footnotesize$\sigma^2_{\Delta_i}=\SI{e-5}{\radian^2}$}] at (19,0.14) {};
\node[label={\footnotesize Common Osci.}] at (28,0.25) {};

\node[ellipse,densely dotted, minimum height=0.6cm,minimum width=0.25cm,draw,label={[anchor=north,xshift=3mm]above:\footnotesize$\sigma^2_{\Delta_i}=\SI{e-4}{\radian^2}$}] at (18,0.325) {};

\end{axis}
\end{tikzpicture}%
\label{fig:EVM_snr}}
\tikzexternaldisable
\hfill
\tikzexternalenable
\tikzsetnextfilename{EVM_Lcomp}
\subfloat[]{\columnplot
%
%

\definecolor{mycolor1}{rgb}{1,0.47058823529,0.47058823529}

\begin{tikzpicture}

\begin{axis}[%
width=\figurewidth,
height=\figureheight,
scale only axis,
xmode=log,
xmin=100,
xmax=10000,
xminorticks=true,
xlabel={$L_f$ in Symbols},
xmajorgrids,
xminorgrids,
ymin=0,
ymax=0.5,
ylabel={EVM},
ymajorgrids,
mark repeat={3},
legend style={at={(0.02,0.98)},anchor=north west,draw=black,fill=white,legend cell align=left,font=\footnotesize}
]
\addplot [color=black,thick,dashed]
  table[row sep=crcr]{100	0.056262602\\
122	0.05605851\\
148	0.056171974\\
178	0.056301737\\
216	0.056049751\\
262	0.056267765\\
318	0.056224226\\
384	0.056288618\\
466	0.056304611\\
564	0.056167661\\
682	0.056252861\\
826	0.056210507\\
1000	0.056177212\\
1212	0.05621217\\
1468	0.056233455\\
1780	0.056236011\\
2156	0.056219614\\
2612	0.05621566\\
3164	0.056226431\\
3832	0.056305232\\
4642	0.056241258\\
5624	0.056250001\\
6814	0.056239742\\
8256	0.056230602\\
10000	0.056255213\\
};
\addlegendentry{No PN};

\addplot [color=blue,solid]
  table[row sep=crcr]{100	0.065426442\\
122	0.066272906\\
148	0.068351426\\
178	0.069629847\\
216	0.072309137\\
262	0.076730467\\
318	0.078113264\\
384	0.082596241\\
466	0.087026125\\
564	0.091708676\\
682	0.09949988\\
826	0.108244469\\
1000	0.114357988\\
1212	0.124795621\\
1468	0.131355129\\
1780	0.147872212\\
2156	0.15331089\\
2612	0.172297266\\
3164	0.181508993\\
3832	0.193765279\\
4642	0.216161303\\
5624	0.240015736\\
6814	0.25744221\\
8256	0.284665856\\
10000	0.302965712\\
};
\addlegendentry{Wiener PN};


\addplot [color=blue,solid,forget plot]
  table[row sep=crcr]{100	0.109480625\\
122	0.120232199\\
148	0.132709171\\
178	0.144495524\\
216	0.154519102\\
262	0.166841947\\
318	0.181986496\\
384	0.192993827\\
466	0.216982114\\
564	0.240696658\\
682	0.2560583\\
826	0.281997224\\
1000	0.303578431\\
1212	0.32676422\\
1468	0.368849298\\
1780	0.416004015\\
2156	0.446117177\\
2612	0.485595322\\
3164	0.533400793\\
3832	0.579807123\\
4642	0.636069304\\
5624	0.67813426\\
6814	0.724847847\\
8256	0.799461902\\
10000	0.883231821\\
};


\addplot [color=mycolor1,solid]
  table[row sep=crcr]{100	0.084311556\\
122	0.085981678\\
148	0.089122363\\
178	0.092409425\\
216	0.096368706\\
262	0.100669892\\
318	0.100843565\\
384	0.109527898\\
466	0.114068489\\
564	0.117948189\\
682	0.124228641\\
826	0.127857752\\
1000	0.135326724\\
1212	0.135121158\\
1468	0.139172851\\
1780	0.140497253\\
2156	0.144837798\\
2612	0.147770159\\
3164	0.148112196\\
3832	0.150082204\\
4642	0.150525966\\
5624	0.151973766\\
6814	0.153377368\\
8256	0.154862177\\
10000	0.155581197\\
};
\addlegendentry{Stationary PN};

\node[label={\footnotesize Reynolds}] at (5*10^3,0.09) {};

\node[label={\footnotesize$\sigma^2_{\Delta_i}=\SI{e-5}{\radian^2}$}] at (1.5*10^3,0.15) {};

\node[label={\footnotesize$\sigma^2_{\Delta_i}=\SI{e-4}{\radian^2}$}] at (3*10^2,0.22) {};

\end{axis}
\end{tikzpicture}%
\label{fig:EVM_length}}
\tikzexternaldisable
\caption{EVM performance of a LOS MIMO system subject to phase noise with $N=4$, $K=\infty$, 16-QAM, perfect channel knowledge: \protect\subref{fig:EVM_snr}~Different oscillator models, $L_f=10^3$, individual setup except dotted curve; \protect\subref{fig:EVM_length}~Impact of frame length, $\text{SNR}=\SI{25}{\decibel}$, individual oscillators.}
\label{fig:EVM}
\end{figure}
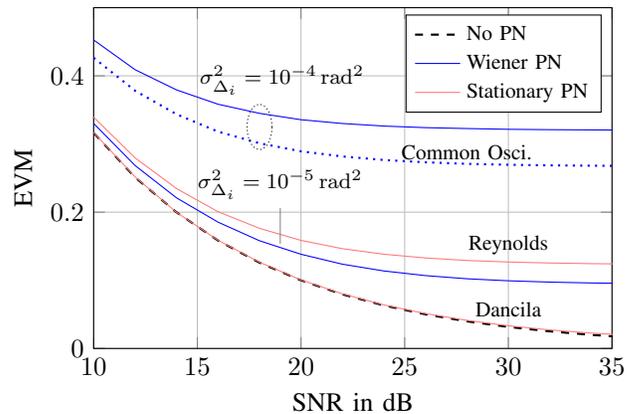
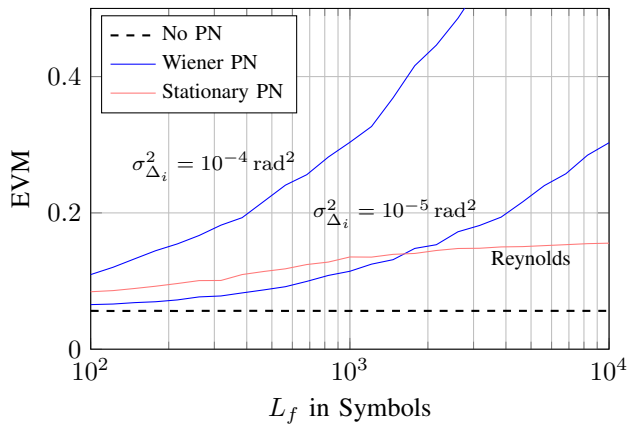

\section{Numerical Results}
\begin{figure*}[!t]
\centering
\tikzexternalenable
\tikzsetnextfilename{SER_overview}
\subfloat[]{\columnplot
%
%
\definecolor{mycolor1}{rgb}{1,0.47058823529,0.47058823529}

\begin{tikzpicture}

\begin{axis}[%
width=\figurewidth,
height=\figureheight,
scale only axis,
xmin=0,
xmax=35,
xlabel={SNR in dB},
xmajorgrids,
ymode=log,
ymin=1e-02,
ymax=1,
yminorticks=true,
ylabel={SER},
ymajorgrids,
yminorgrids,
legend style={at={(0.02,0.02)},anchor=south west,draw=black,fill=white,legend cell align=left,font=\footnotesize}
]

\addplot [color=black,thick,dashed]
  table[row sep=crcr]{-5	0.8498225\\
-3	0.8216475\\
-1	0.7825075\\
1	0.7326375\\
3	0.6642\\
5	0.57447\\
7	0.4658875\\
9	0.3397925\\
11	0.21232\\
13	0.1053475\\
15	0.0378975\\
17	0.0087775\\
19	0.00112\\
21	7e-05\\
23	2.5e-06\\
25	0\\
27	0\\
29	0\\
31	0\\
33	0\\
35	0\\
37	0\\
39	0\\
};
\addlegendentry{No PN};

\addplot [color=blue,solid,dotted,thick,forget plot]
  table[row sep=crcr]{-5	0.8559375\\
-3	0.8322725\\
-1	0.800885\\
1	0.7577175\\
3	0.7046875\\
5	0.63886\\
7	0.559335\\
9	0.473895\\
11	0.389785\\
13	0.3171275\\
15	0.2656275\\
17	0.2326575\\
19	0.2141525\\
21	0.205995\\
23	0.200165\\
25	0.19727\\
27	0.195875\\
29	0.1945975\\
31	0.1942675\\
33	0.1935975\\
35	0.1934\\
37	0.19346\\
39	0.19331\\
};

\addplot [color=blue,solid]
  table[row sep=crcr]{-5	0.8582325\\
-3	0.8333175\\
-1	0.801395\\
1	0.7609325\\
3	0.71022\\
5	0.6487925\\
7	0.574985\\
9	0.49788\\
11	0.42328\\
13	0.3588925\\
15	0.3101375\\
17	0.278715\\
19	0.2574925\\
21	0.2447425\\
23	0.23735\\
25	0.2327675\\
27	0.2297425\\
29	0.22797\\
31	0.226965\\
33	0.226465\\
35	0.2259825\\
37	0.22602\\
39	0.22571\\
};
\addlegendentry{Wiener PN};

\addplot [color=blue,solid,forget plot]
  table[row sep=crcr]{-5	0.8582975\\
-3	0.834395\\
-1	0.80293625\\
1	0.76287625\\
3	0.71293375\\
5	0.651215\\
7	0.5814325\\
9	0.5081725\\
11	0.43769375\\
13	0.37690625\\
15	0.331358125\\
17	0.300185\\
19	0.280143125\\
21	0.267984375\\
23	0.260664375\\
25	0.255655\\
27	0.2529525\\
29	0.251061875\\
31	0.249954375\\
33	0.24915\\
35	0.24886625\\
37	0.248535625\\
39	0.24837125\\
};

\addplot [color=blue,solid,forget plot]
  table[row sep=crcr]{-5	0.859311198\\
-3	0.835953385\\
-1	0.804908333\\
1	0.765873438\\
3	0.716201563\\
5	0.657421875\\
7	0.590292708\\
9	0.520072656\\
11	0.4530375\\
13	0.39565625\\
15	0.351532031\\
17	0.320996615\\
19	0.301002604\\
21	0.288214583\\
23	0.280118229\\
25	0.275120573\\
27	0.27177474\\
29	0.2698875\\
31	0.268651042\\
33	0.267793229\\
35	0.267432813\\
37	0.267013021\\
39	0.266761719\\
};

\addplot [color=mycolor1,solid]
  table[row sep=crcr]{-5	0.845285\\
-3	0.8150925\\
-1	0.7751925\\
1	0.7222625\\
3	0.6561325\\
5	0.5715175\\
7	0.4716475\\
9	0.361225\\
11	0.2580375\\
13	0.1715175\\
15	0.113575\\
17	0.07777\\
19	0.058465\\
21	0.04818\\
23	0.0425925\\
25	0.038715\\
27	0.03661\\
29	0.0355025\\
31	0.0346975\\
33	0.03444\\
35	0.0338925\\
37	0.033725\\
39	0.033735\\
};
\addlegendentry{Stationary PN};

\node[label={\footnotesize Reynolds, $N=4$}] at (28,0.018) {};

\node[ellipse,densely dotted, minimum height=0.6cm,minimum width=0.25cm,draw,label={[anchor=north,xshift=3mm]above:\footnotesize$\sigma^2_{\Delta_i}=\SI{e-4}{\radian^2}$}] at (15,0.325) {};

\node[pin=250:{\footnotesize$N=4$}] at (22.5,0.27) {};
\node[pin=270:{\footnotesize$N=16$}] at (24.5,0.29) {};
\node[pin=290:{\footnotesize$N=96$}] at (26.5,0.30) {};

\node[pin=250:{\footnotesize Common Osci.}] at (12.5,0.37) {};
\node[label={\footnotesize $N=4$}] at (5.5,0.085) {};

\end{axis}
\end{tikzpicture}%
\label{fig:SER1}}
\tikzexternaldisable
\hfill
\tikzexternalenable
\tikzsetnextfilename{SER_constellations}
\subfloat[]{\columnplot
%
%
\begin{tikzpicture}

\begin{axis}[%
width=\figurewidth,
height=\figureheight,
scale only axis,
xmin=0,
xmax=35,
xlabel={SNR in dB},
xmajorgrids,
ymode=log,
ymin=1e-04,
ymax=1,
yminorticks=true,
ylabel={SER},
ymajorgrids,
yminorgrids,
mark repeat=3,
mark options={solid},
legend style={at={(0.02,0.02)},anchor=south west,draw=black,fill=white,legend cell align=left,font=\footnotesize}
]

\addplot [color=blue,solid,forget plot,dashed]
  table[row sep=crcr]{-5	0.848135\\
-3	0.8198575\\
-1	0.78188\\
1	0.729285\\
3	0.6620325\\
5	0.5738975\\
7	0.46846\\
9	0.3492475\\
11	0.2304525\\
13	0.12972\\
15	0.06309\\
17	0.0279125\\
19	0.0125025\\
21	0.0065275\\
23	0.0041925\\
25	0.00307\\
27	0.002355\\
29	0.0020725\\
31	0.00194\\
33	0.001875\\
35	0.0017325\\
37	0.001745\\
39	0.0017725\\
};

\addplot [color=blue,solid,forget plot,dashed,mark=square]
  table[row sep=crcr]{-5	0.9583625\\
-3	0.94985\\
-1	0.93833\\
1	0.921685\\
3	0.8984975\\
5	0.8673825\\
7	0.82228\\
9	0.7624125\\
11	0.68433\\
13	0.5868975\\
15	0.4792075\\
17	0.3672625\\
19	0.2692775\\
21	0.1946575\\
23	0.147115\\
25	0.1194775\\
27	0.103455\\
29	0.09518\\
31	0.08953\\
33	0.08674\\
35	0.0845225\\
37	0.083315\\
39	0.082645\\
};

\addplot [color=blue,solid,forget plot,dashed,mark=o]
  table[row sep=crcr]{-5	0.7365975\\
-3	0.69234\\
-1	0.6407225\\
1	0.5687925\\
3	0.4845425\\
5	0.38435\\
7	0.2834375\\
9	0.185875\\
11	0.1056175\\
13	0.052255\\
15	0.0222\\
17	0.009005\\
19	0.0035325\\
21	0.0015325\\
23	0.0007575\\
25	0.0004725\\
27	0.0002875\\
29	0.0001875\\
31	0.000125\\
33	7.5e-05\\
35	8.75e-05\\
37	7.75e-05\\
39	9e-05\\
};

\addplot [color=blue,solid,forget plot,dashed,mark=star]
  table[row sep=crcr]{-5	0.8649375\\
-3	0.8419\\
-1	0.811385\\
1	0.7716925\\
3	0.7192525\\
5	0.6570925\\
7	0.581335\\
9	0.4980825\\
11	0.40854\\
13	0.316905\\
15	0.2369625\\
17	0.170695\\
19	0.122485\\
21	0.092735\\
23	0.074485\\
25	0.06249\\
27	0.0562125\\
29	0.0523775\\
31	0.0494925\\
33	0.04812\\
35	0.04702\\
37	0.04663\\
39	0.0460075\\
};

\addplot [color=blue,solid]
  table[row sep=crcr]{-5	0.8582325\\
-3	0.8333175\\
-1	0.801395\\
1	0.7609325\\
3	0.71022\\
5	0.6487925\\
7	0.574985\\
9	0.49788\\
11	0.42328\\
13	0.3588925\\
15	0.3101375\\
17	0.278715\\
19	0.2574925\\
21	0.2447425\\
23	0.23735\\
25	0.2327675\\
27	0.2297425\\
29	0.22797\\
31	0.226965\\
33	0.226465\\
35	0.2259825\\
37	0.22602\\
39	0.22571\\
};
\addlegendentry{16-QAM};

\addplot [color=blue,solid,mark=square]
  table[row sep=crcr]{-5	0.961\\
-3	0.95419\\
-1	0.9445175\\
1	0.9317025\\
3	0.9149025\\
5	0.89343\\
7	0.866405\\
9	0.8330825\\
11	0.797215\\
13	0.754155\\
15	0.71224\\
17	0.6746325\\
19	0.6427275\\
21	0.61764\\
23	0.5996875\\
25	0.58928\\
27	0.5821475\\
29	0.57729\\
31	0.5743925\\
33	0.57247\\
35	0.5712625\\
37	0.5705725\\
39	0.570345\\
};
\addlegendentry{64-QAM};

\addplot [color=blue,solid,mark=o]
  table[row sep=crcr]{-5	0.74815\\
-3	0.71117\\
-1	0.666405\\
1	0.60888\\
3	0.54372\\
5	0.471615\\
7	0.403025\\
9	0.34084\\
11	0.2898525\\
13	0.2536425\\
15	0.229925\\
17	0.214305\\
19	0.204575\\
21	0.1984\\
23	0.194055\\
25	0.1920075\\
27	0.19045\\
29	0.18947\\
31	0.1887875\\
33	0.1886225\\
35	0.1881075\\
37	0.1881475\\
39	0.1881125\\
};
\addlegendentry{8-PSK};

\addplot [color=blue,solid,mark=star]
  table[row sep=crcr]{-5	0.8714775\\
-3	0.8509125\\
-1	0.825145\\
1	0.7930225\\
3	0.7545725\\
5	0.7105075\\
7	0.6627875\\
9	0.6178225\\
11	0.5746625\\
13	0.536585\\
15	0.50655\\
17	0.484765\\
19	0.4680525\\
21	0.4579125\\
23	0.4509175\\
25	0.4461375\\
27	0.443445\\
29	0.4418675\\
31	0.44062\\
33	0.4400875\\
35	0.4396925\\
37	0.4392075\\
39	0.438905\\
};
\addlegendentry{16-PSK};

\node[rectangle,draw=blue] at (7,0.015) {\footnotesize$\sigma^2_{\Delta_i}=\SI{e-4}{\radian^2}$};

\node[rectangle,draw=blue,dashed] at (28,0.015) {\footnotesize$\sigma^2_{\Delta_i}=\SI{e-5}{\radian^2}$};

\end{axis}
\end{tikzpicture}%
\label{fig:SER2}}
\tikzexternaldisable
\caption{SER performance of a LOS MIMO subject to phase noise with $K=\SI{10}{\decibel}$: \protect\subref{fig:SER1}~Different phase noise models, individual setup except dotted curve; \protect\subref{fig:SER2}~Different modulation schemes, $N=4$, Wiener PN (solid $\sigma^2_{\Delta_i}=\SI{e-4}{\radian^2}$, dashed $\sigma^2_{\Delta_i}=\SI{e-5}{\radian^2}$), individual oscillator setup.}
\label{fig:SER}
\end{figure*}
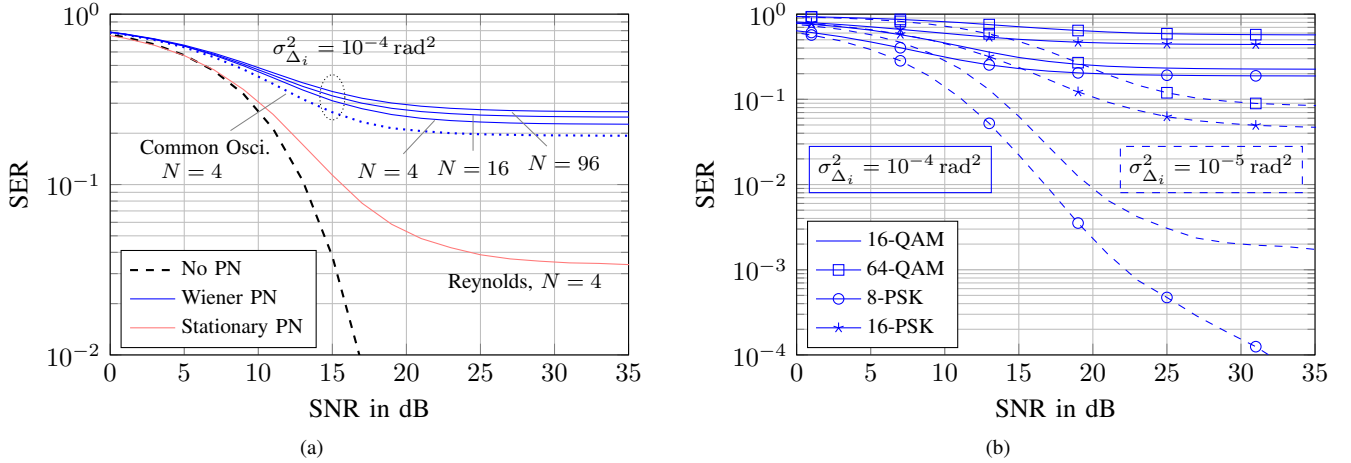

In this section we will show the impact of phase noise for different LOS MIMO system designs. For fair comparison we will apply a transmit power constraint with $\sigma^2_w = \frac{1}{\text{SNR}_\text{Lin}/N}$ where $\text{SNR}=10\log_{10} (\text{SNR}_\text{Lin})$. Furthermore, we will set the initial phases of the Wiener model $\theta_i(0)$ and the initial phases of the stationary model $\theta_{i,0}$ to zero. In practice it will occur, particularly for the individual oscillator setup, that there is a fixed phase offset between the different oscillators and antennas. However, as this does not change the conditioning of the channel matrix, it will not influence the performance and is, for example, compensated by the initial channel estimation. We also presume that the same oscillators are used at Tx and Rx, i.e., same phase noise variances $\sigma^2_{\Delta_i}$ for Wiener PN and same power spectral densities for stationary PN.

\subsection{Error Vector Magnitude}
First, we show the impact of phase noise on the error~vector~magnitude~(EVM) under perfect channel knowledge $\hat{\matr{H}}=\matr{H}$ in a pure LOS channel, i.e., $K=\infty$. The EVM will be defined as 
\begin{equation}
\text{EVM} = \expect\left[\sqrt{\frac{ \left|\matr{x}(n)-\hat{\matr{x}}(n)\right|^2}{E_x}}\right]
\end{equation}
where $E_x$ is the average symbol energy.

In Fig.~\ref{fig:EVM} we show the EVM performance for different oscillator setups when using 16-QAM. It can be seen that the influence from stationary phase noise is much less severe compared to the impact of Wiener PN. For high quality oscillators, as proposed in \cite{Dancila2014}, there is almost no degradation observable. Secondly, for higher phase noise values the EVM does not tend towards zero with increasing SNR but runs towards a fixed value, which means that in the high SNR domain phase noise will be the system performance limiting parameter. As expected, the EVM of the common oscillator setup outperforms the individual oscillator setup. However, the gap is rather small, scales insignificantly with the number of antennas $N$ (not shown here) and will be less for lower phase noise values. Fig.~\ref{fig:EVM_length} shows the performance against frame length $L_f$ at an SNR of \SI{25}{\decibel}. It can be seen that for stationary phase noise the EVM is almost independent of the frame length, since the phase is locked and hence only small variations can occur. For Wiener PN on the other hand the frame length will have a significant impact on the performance if no additional synchronization is performed. This is due to the fact that only the frequency is locked and therefore the phase will eventually run away.

\subsection{Symbol Error Rate}
Next we will investigate the symbol~error~rate~(SER) performance of different scenarios and show the efficiency of the compensation algorithm. 
For a more realistic scenario, $\hat{\matr{H}}$ will be gained from a BPSK Hadamard training sequence of length $L_t=N$ and we set $K=\SI{10}{\decibel}$. Based on the observations made in the previous section, $L_d$ will be chosen to be $10^3$ and the phase will be synchronized again between frames, so that the variance of the Wiener process does not grow boundlessly. The symbol estimates $\bar{x}_i$ will be gained by performing ML~detection on the symbols $\hat{x}_i$.

Fig.~\ref{fig:SER} shows the SER of different system setups. As for the EVM it can also be seen here that the difference between common and individual oscillator setup is small. Furthermore, the results also show that the impact of phase noise scales only marginally with the number of antennas in terms of error rate, and that the impact of stationary PN is less severe. In Fig.~\ref{fig:SER2} the sensitivity of different modulation formats in a system with $N=4$ and individual oscillator setup is displayed for Wiener phase noise. As expected PSK constellations are more effected by PN than QAM constellations.

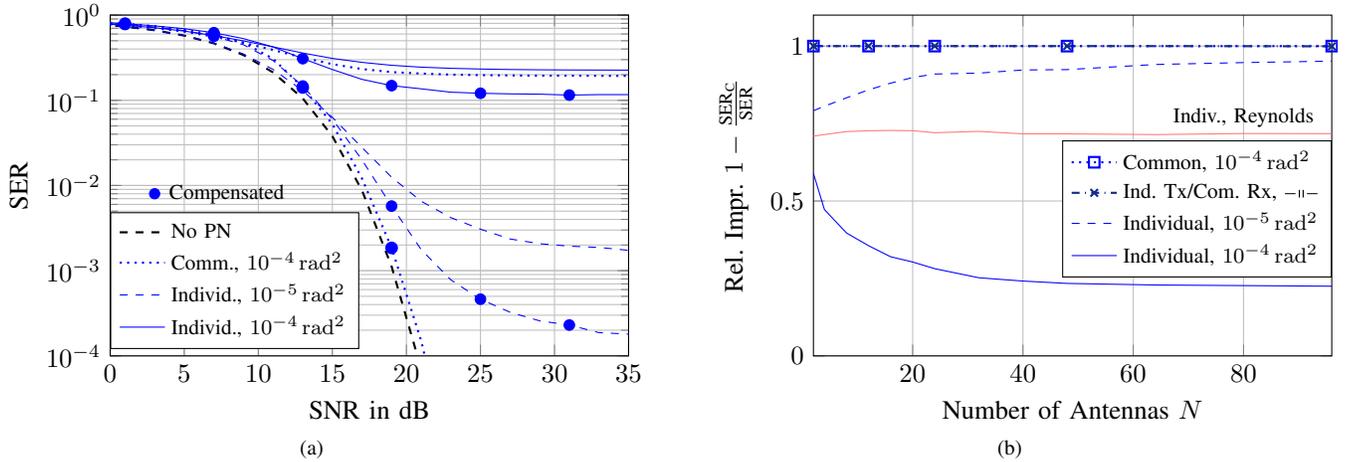
\begin{figure*}[!t]
\centering
\tikzexternalenable
\tikzsetnextfilename{SER_compensated}
\subfloat[]{\columnplot
%
%
\begin{tikzpicture}

\begin{axis}[%
width=\figurewidth,
height=\figureheight,
scale only axis,
xmin=0,
xmax=35,
xlabel={SNR in dB},
xmajorgrids,
ymode=log,
ymin=1e-04,
ymax=1,
yminorticks=true,
ylabel={$\text{SER}$},
ymajorgrids,
yminorgrids,
mark repeat=3,
mark options={solid},
legend style={at={(0.00,0.02)},anchor=south west,draw=black,fill=white,legend cell align=left,font=\footnotesize}
]

\addplot [color=black,solid,dashed,thick]
  table[row sep=crcr]{-5	0.8498225\\
-3	0.8216475\\
-1	0.7825075\\
1	0.7326375\\
3	0.6642\\
5	0.57447\\
7	0.4658875\\
9	0.3397925\\
11	0.21232\\
13	0.1053475\\
15	0.0378975\\
17	0.0087775\\
19	0.00112\\
21	7e-05\\
23	2.5e-06\\
25	0\\
27	0\\
29	0\\
31	0\\
33	0\\
35	0\\
37	0\\
39	0\\
};
\addlegendentry{No PN};

\addplot [color=blue,solid,dotted,thick]
  table[row sep=crcr]{-5	0.8559375\\
-3	0.8322725\\
-1	0.800885\\
1	0.7577175\\
3	0.7046875\\
5	0.63886\\
7	0.559335\\
9	0.473895\\
11	0.389785\\
13	0.3171275\\
15	0.2656275\\
17	0.2326575\\
19	0.2141525\\
21	0.205995\\
23	0.200165\\
25	0.19727\\
27	0.195875\\
29	0.1945975\\
31	0.1942675\\
33	0.1935975\\
35	0.1934\\
37	0.19346\\
39	0.19331\\
};
\addlegendentry{Comm., \SI{e-4}{\radian^2}};

\addplot [color=blue,solid,forget plot,thick,dotted,mark=*]
  table[row sep=crcr]{-5	0.874105\\
-3	0.856755\\
-1	0.8268375\\
1	0.7923175\\
3	0.74184\\
5	0.671755\\
7	0.58874\\
9	0.474465\\
11	0.3071425\\
13	0.141295\\
15	0.0526725\\
17	0.012085\\
19	0.00185\\
21	0.00015\\
23	5e-06\\
25	0\\
27	0\\
29	0\\
31	0\\
33	0\\
35	0\\
37	0\\
39	0\\
};

\addplot [color=blue,solid,dashed]
  table[row sep=crcr]{-5	0.848135\\
-3	0.8198575\\
-1	0.78188\\
1	0.729285\\
3	0.6620325\\
5	0.5738975\\
7	0.46846\\
9	0.3492475\\
11	0.2304525\\
13	0.12972\\
15	0.06309\\
17	0.0279125\\
19	0.0125025\\
21	0.0065275\\
23	0.0041925\\
25	0.00307\\
27	0.002355\\
29	0.0020725\\
31	0.00194\\
33	0.001875\\
35	0.0017325\\
37	0.001745\\
39	0.0017725\\
};
\addlegendentry{Individ., \SI{e-5}{\radian^2}};

\addplot [color=blue,solid,forget plot,dashed,mark=*]
  table[row sep=crcr]{-5	0.870675\\
-3	0.848905\\
-1	0.816205\\
1	0.7771775\\
3	0.7246425\\
5	0.657215\\
7	0.563795\\
9	0.4484775\\
11	0.29244\\
13	0.1469575\\
15	0.0602425\\
17	0.019985\\
19	0.0057275\\
21	0.001765\\
23	0.00078\\
25	0.0004625\\
27	0.000325\\
29	0.0002575\\
31	0.00023\\
33	0.0001875\\
35	0.00018\\
37	0.0001725\\
39	0.00017\\
};

\addplot [color=blue,solid]
  table[row sep=crcr]{-5	0.8582325\\
-3	0.8333175\\
-1	0.801395\\
1	0.7609325\\
3	0.71022\\
5	0.6487925\\
7	0.574985\\
9	0.49788\\
11	0.42328\\
13	0.3588925\\
15	0.3101375\\
17	0.278715\\
19	0.2574925\\
21	0.2447425\\
23	0.23735\\
25	0.2327675\\
27	0.2297425\\
29	0.22797\\
31	0.226965\\
33	0.226465\\
35	0.2259825\\
37	0.22602\\
39	0.22571\\
};
\addlegendentry{Individ., \SI{e-4}{\radian^2}};

\addplot [color=blue,solid,forget plot,mark=*]
  table[row sep=crcr]{-5	0.8731525\\
-3	0.8536225\\
-1	0.83104\\
1	0.7927625\\
3	0.74717\\
5	0.69245\\
7	0.6211325\\
9	0.5292775\\
11	0.4229675\\
13	0.308245\\
15	0.229685\\
17	0.1760275\\
19	0.14861\\
21	0.1363125\\
23	0.1244725\\
25	0.120865\\
27	0.11891\\
29	0.11798\\
31	0.1150925\\
33	0.1165375\\
35	0.11655\\
37	0.11385\\
39	0.11518\\
};

\node[circle,draw=blue,fill=blue,scale=0.41,label={[anchor=west,xshift=0mm]right:\footnotesize Compensated}] at (3,0.008) {};

\end{axis}
\end{tikzpicture}%
\label{fig:SERc1}}
\tikzexternaldisable
\hfill
\tikzexternalenable
\tikzsetnextfilename{SER_r_ant}
\subfloat[]{\columnplot
%
%

\definecolor{mycolor1}{rgb}{1,0.47058823529,0.47058823529}
\definecolor{mycolor2}{rgb}{0.0784313753247261,0.168627455830574,0.549019634723663}%

\begin{tikzpicture}

\begin{axis}[%
width=\figurewidth,
height=\figureheight,
scale only axis,
xmin=2,
xmax=96,
xlabel={Number of Antennas $N$},
xmajorgrids,
ymin=0.0,
ymax=1.1,
ylabel={Rel. Impr. $1-\frac{\text{SER}_\text{C}}{\text{SER}}$},
ymajorgrids,
mark repeat=3,
mark options={solid},
legend style={at={(1.0,0.23)},anchor=south east,draw=black,fill=white,legend cell align=left,font=\footnotesize}
]

\addplot [color=blue,solid,dotted,thick,mark=square]
  table[row sep=crcr]{2	1\\
4	1\\
8	1\\
12	1\\
16	1\\
20	1\\
24	1\\
32	0.99999386\\
40	1\\
48	0.99998483\\
64	0.999947884\\
80	0.999931514\\
96	0.999942749\\
};
\addlegendentry{Common, \SI{e-4}{\radian^2}};

\addplot [color=mycolor2,solid, dash dot,mark=x,thick]
  table[row sep=crcr]{2	1\\
4	1\\
8	1\\
12	1\\
16	1\\
20	1\\
24	1\\
32	1\\
40	0.999983444\\
48	0.999989969\\
64	0.999973887\\
80	0.999735611\\
96	0.999246568\\
};
\addlegendentry{Ind. Tx/Com. Rx, \tikz{
        \draw [line width=0.12ex] (-0.2ex,0) -- +(0,0.8ex)
																	(0.2ex,0) -- +(0,0.8ex);
        \draw [line width=0.08ex] (-0.6ex,0.4ex) -- +(-0.5em,0)
            											(0.6ex,0.4ex) -- +(0.5em,0);
    }};

\addplot [color=blue,solid,dashed]
  table[row sep=crcr]{2	0.791613043\\
4	0.805930979\\
8	0.8330752\\
12	0.858809677\\
16	0.880392241\\
20	0.89860041\\
24	0.909881748\\
32	0.9128\\
40	0.922933936\\
48	0.924120486\\
64	0.940680311\\
80	0.946788695\\
96	0.951440401\\
};
\addlegendentry{Individual, \SI{e-5}{\radian^2}}; 

\addplot [color=blue,solid]
  table[row sep=crcr]{2	0.588913713\\
4	0.472226887\\
8	0.396781004\\
12	0.355718994\\
16	0.320021668\\
20	0.303022323\\
24	0.281482907\\
32	0.2522296\\
40	0.241563976\\
48	0.233947768\\
64	0.228853451\\
80	0.226889999\\
96	0.224942858\\
};
\addlegendentry{Individual, \SI{e-4}{\radian^2}};

\addplot [color=mycolor1,solid,forget plot]
  table[row sep=crcr]{2	0.70936094\\
4	0.714388724\\
8	0.723971331\\
12	0.726765954\\
16	0.727692386\\
20	0.726924132\\
24	0.720403085\\
32	0.724732708\\
40	0.716733104\\
48	0.716644159\\
64	0.714378781\\
80	0.717736878\\
96	0.71731828\\
};


\node[label={\footnotesize Indiv., Reynolds}] at (80,0.68) {};

\end{axis}
\end{tikzpicture}%
\label{fig:SERc2}}
\tikzexternaldisable
\caption{Performance of a LOS MIMO system compensated by the shown algorithm with $K=\SI{10}{\decibel}$, 16-QAM: \protect\subref{fig:SERc1}~SER for $N=4$, different stetups with Wiener PN; \protect\subref{fig:SERc2}~Relative SER improvement of the algorithm for $\text{SNR}=\SI{25}{\decibel}$ and different oscillator setups.}
\label{fig:SER_comp}
\end{figure*}

In Fig.~\ref{fig:SERc1} the performance of the proposed compensation algorithm is evaluated. We set $\alpha=0.1$, as this showed the best performance over a wide range of scenarios. This corresponds to a slow convergence, i.e., the algorithm will not be able to follow fast phase changes. First, for the common oscillator setup the algorithm can improve the performance of the system significantly, even for a high phase noise variance of $\sigma^2_{\Delta_i}=\SI{e-4}{\radian^2}$. For the individual oscillator setup there is also an improvement in terms of SER, but especially for high PN values the error floor is still significant.

Finally, Fig.~\ref{fig:SERc2} shows the relative improvement in terms of SER that the algorithm achieves for different numbers of antennas at an SNR of \SI{25}{\decibel}. Noteworthy are the two cases where the relative improvement is $1$, i.e., the compensated symbol error rate $\text{SER}_\text{C}$ is zero. This happens for I.) common oscillator at transmitter and receiver, and II.) individual oscillators at Tx and a common oscillator at Rx. The second structure works well because it effectively exploits the same property that the common oscillator uses. It can be shown that for this case, when using a ZF receiver, the total phase noise is also just an addition of the corresponding two processes from transmitter and receiver. Overall, the influence of the antenna number $N$ on the algorithm is not substantial.

\section{Conclusion}

In this paper a communications system based on LOS MIMO has been investigated regarding its sensitivity towards phase noise. Two different phase noise models were introduced and their impact in terms of EVM was studied for practical values. The results show that phase noise that follows a Wiener process is a lot more critical than stationary phase noise. Furthermore, the difference between a common and individual oscillator setup is not immense regarding raw error magnitude. 

Additionally, we demonstrated a phase noise compensation scheme and investigated its system level performance. The algorithm shows good behavior for the case of a common oscillator setup, even for high phase noise values. This is intuitive as one only has to track one phase noise process but has $N$ observations of the process. For individual oscillators at transmitter and receiver the algorithm generates less performance gain, but shows for low PN values still considerable improvements. To combat higher phase noises in this setup it seems unavoidable to use more complex algorithms, such as higher order feedback loops or the extended Kalman filter proposed in \cite{Mehrpouyan2012}. However, this algorithm also requires additional pilots and is far more complex. Ultimately, it will be necessary to find the best trade-off between required phase noise compensation and its implementation complexity for a desired system performance.

\IEEEtriggeratref{2}
\IEEEtriggercmd{\enlargethispage{2mm}}


\bibliographystyle{IEEEtran}
\bibliography{references}

\begin{thebibliography}{10}
\providecommand{\url}[1]{#1}
\csname url@samestyle\endcsname
\providecommand{\newblock}{\relax}
\providecommand{\bibinfo}[2]{#2}
\providecommand{\BIBentrySTDinterwordspacing}{\spaceskip=0pt\relax}
\providecommand{\BIBentryALTinterwordstretchfactor}{4}
\providecommand{\BIBentryALTinterwordspacing}{\spaceskip=\fontdimen2\font plus
\BIBentryALTinterwordstretchfactor\fontdimen3\font minus
  \fontdimen4\font\relax}
\providecommand{\BIBforeignlanguage}[2]{{%
\expandafter\ifx\csname l@#1\endcsname\relax
\typeout{** WARNING: IEEEtran.bst: No hyphenation pattern has been}%
\typeout{** loaded for the language `#1'. Using the pattern for}%
\typeout{** the default language instead.}%
\else
\language=\csname l@#1\endcsname
\fi
#2}}
\providecommand{\BIBdecl}{\relax}
\BIBdecl

\bibitem{Bohagen2007}
F.~B{\o}hagen, P.~Orten, and G.~E. {\O}ien, ``{Design of Optimal High-Rank
  Line-of-Sight MIMO Channels},'' \emph{IEEE Trans. Wirel. Commun.}, vol.~6,
  no.~4, pp. 1420--1425, 2007.

\bibitem{Sheldon2009}
C.~Sheldon, M.~Seo, E.~Torkildson, M.~Rodwell, and U.~Madhow, ``{Four-Channel
  Spatial Multiplexing Over a Millimeter-Wave Line-of-Sight Link},'' in
  \emph{Proc. Int. Microw. Symp.}, 2009, pp. 389--392.

\bibitem{Schenk2008}
T.~Schenk, \emph{{RF Imperfections in High-rate Wireless Systems}}.\hskip 1em
  plus 0.5em minus 0.4em\relax Springer, 2008.

\bibitem{Bjornson2013}
E.~Bj\"{o}rnson, P.~Zetterberg, M.~Bengtsson, and B.~Ottersten, ``{Capacity
  Limits and Multiplexing Gains of MIMO Channels with Transceiver
  Impairments},'' \emph{IEEE Commun. Lett.}, vol.~17, no.~1, pp. 91--94, 2013.

\bibitem{Zhang2014}
X.~Zhang, M.~Matthaiou, E.~Bj\"{o}rnson, M.~Coldrey, and M.~Debbah, ``{On the
  MIMO Capacity with Residual Transceiver Hardware Impairments},'' in
  \emph{IEEE Int. Conf. Commun.}, 2014, pp. 5299--5305.

\bibitem{Durisi2013}
G.~Durisi, A.~Tarable, and T.~Koch, ``{On the Multiplexing Gain of MIMO
  Microwave Backhaul Links Affected by Phase Noise},'' in \emph{Proc. IEEE Int.
  Conf. Commun.}, 2013, pp. 3209--3214.

\bibitem{Durisi2014}
G.~Durisi, A.~Tarable, C.~Camarda, R.~Devassy, and G.~Montorsi, ``{Capacity
  Bounds for MIMO Microwave Backhaul Links Affected by Phase Noise},''
  \emph{IEEE Trans. Commun.}, vol.~62, no.~3, pp. 920--929, 2014.

\bibitem{Mehrpouyan2012}
H.~Mehrpouyan, A.~A. Nasir, S.~D. Blostein, T.~Eriksson, G.~K. Karagiannidis,
  and T.~Svensson, ``{Joint Estimation of Channel and Oscillator Phase Noise in
  MIMO Systems},'' \emph{IEEE Trans. Signal Process.}, vol.~60, no.~9, pp.
  4790--4807, 2012.

\bibitem{Muhi-Eldeen2010}
Z.~Muhi-Eldeen, L.~Ivrissimtzis, and M.~Al-Nuaimi, ``{Modelling and
  measurements of millimetre wavelength propagation in urban environments},''
  \emph{IET Microwaves, Antennas Propag.}, vol.~4, no.~9, pp. 1300--1309, 2010.

\bibitem{Khanzadi2014}
M.~R. Khanzadi, D.~Kuylenstierna, A.~Panahi, T.~Eriksson, and H.~Zirath,
  ``{Calculation of the Performance of Communication Systems From Measured
  Oscillator Phase Noise},'' \emph{IEEE Trans. Circuits Syst. I Regul. Pap.},
  vol.~61, no.~5, pp. 1553--1565, 2014.

\bibitem{Demir2000}
A.~Demir, A.~Mehrotra, and J.~Roychowdhury, ``{Phase Noise in Oscillators: A
  Unifying Theory and Numerical Methods for Characterization},'' \emph{IEEE
  Trans. Circuits Syst. I Fundam. Theory Appl.}, vol.~47, no.~5, pp. 655--674,
  May 2000.

\bibitem{Rappaport2011}
T.~S. Rappaport, J.~N. Murdock, and F.~Gutierrez, ``{State of the Art in 60-GHz
  Integrated Circuits and Systems for Wireless Communications},'' \emph{Proc.
  IEEE}, vol.~99, no.~8, pp. 1390--1436, 2011.

\bibitem{Kasdin1995}
N.~J. Kasdin, ``{Discrete Simulation of Colored Noise and Stochastic Processes
  and 1/$f^\alpha$ Power Law Noise Generation},'' \emph{Proc. IEEE}, vol.~83,
  no.~5, pp. 802--827, 1995.

\bibitem{Reynolds2006}
S.~K. Reynolds, B.~A. Floyd, U.~R. Pfeiffer, T.~Beukema, J.~Grzyb, C.~Haymes,
  B.~Gaucher, and M.~Soyuer, ``{A Silicon 60-GHz Receiver and Transmitter
  Chipset for Broadband Communications},'' \emph{IEEE J. Solid-State Circuits},
  vol.~41, no.~12, pp. 2820--2831, 2006.

\bibitem{Dancila2014}
D.~Dancila, X.~Rottenberg, H.~A.~C. Tilmans, W.~{De Raedt}, and I.~Huynen,
  ``{Low Phase Noise Oscillator at 60 GHz Stabilized by a Substrate Integrated
  Cavity Resonator in LTCC},'' \emph{IEEE Microw. Wirel. Components Lett.},
  vol.~24, no.~12, pp. 887--889, Dec. 2014.

\end{thebibliography}
%

\end{document}